# The large-scale energetic ion layer in the high latitude Jovian magnetosphere as revealed by Ulysses/HI-SCALE cross-field intensity-gradient measurements

Anagnostopoulos G., I. Karanikola, P.K. Marhavilas, and E. Sarris

*Space Science Laboratory, Department of Electrical and Computer Engineering,*

*Democritus University of Thrace, 67100 Xanthi, Greece*

**Abstract.** Ulysses investigated the high latitude Jovian magnetosphere for a second time, after Pioneer 11 mission, and gave us the opportunity to search the structure and the dynamics of this giant magnetosphere above the magnetodisc. Kivelson (1976) and Kennel and Coroniti (1979) reported that Pioneer 11 observed energetic particle intensities at high latitudes at the same level with those measured in the plasma sheet and inferred that they were not consistent with the magnetodisc model. Ulysses observations supported the idea about a large-scale layer of energetic ions and electrons in the outer high latitude Jovian magnetosphere (Cowley et al.,1996; Anagnostopoulos et al., 2001). This study perform a number of further tests for the existence of the large scale layer of energetic ions in the outer high latitude Jovian magnetosphere by studying appropriate cross-B field anisotropies in order to monitor the ion northward/southward intensity gradients. In particular, we examined Ulysses/HI-SCALE observations of energetic ions with large gyro-radius (0.5-1.6 MeV protons and >2.5 MeV heavy (Z>5) ions) in order to compare instant intensity changes with remote sensing intensity gradients. Our analysis confirms the existence of an energetic particle layer in the north hemisphere, during the inbound trajectory of Ulysses traveling at moderate latitudes, and in the south high-latitude duskside magnetosphere, during the outbound segment of the spacecraft trajectory. Our Ulysses/HI-SCALE data analysis also provides evidence for the detection of an energetic proton magnetopause boundary layer during the outbound trajectory of the spacecraft. During Ulysses flyby of Jupiter the almost permanent appearance of alternative northward and southward intensity gradients suggests that the high latitude layer appeared to be a third major area of energetic particles, which coexisted with the radiation belts and the magnetodisc.

# 1. Introduction

Magnetic field observations from the Pioneer-10 and 11 and the Voyager-1 and 2 missions, had provided the basis for the identification of three general regions of the Jovian magnetosphere, based mostly on magnetic field measurements: the outer magnetosphere, the middle magnetosphere, and the inner one (Smith et al., 1976; Acuňa et al., 1983, Hawkins et al., 1998). The inner magnetosphere is dominated by the planetary dipole, lies within ~15 planetary radii ($R_J$) and is populated by the radiation belt particles. The middle magnetosphere is dominated by the equatorial current sheet and the magnetic field lines tend to be radial and oppositely directed in the two sides of the "magnetodisc. The outer magnetosphere extends from the end of the magnetodisc to the magnetopause.

Ulysses explored the huge and amazing plasma laboratory of the Jovian magnetosphere for the fifth time, during its voyage to the solar poles, in February of year 1992, and extended our knowledge on the structure and the dynamics of the Jovian magnetosphere gained by the discoveries and the results of the four previous missions (Pioneer-10 and 11, Voyager-1 and 2). An issue that arose during Ulysses' flyby of Jupiter and appears to change the picture that we had about the structure of the Jovian magnetosphere, is the existence of a new discovered layer of energetic particles at high latitudes near the magnetopause [Cowley et al.,1996; Anagnostopoulos et al., 1998; 2001; Krupp et al., 1999; Marhavilas et al. 2004]. This layer arrears to be a third major area of energetic particles, which coexists with the radiation belts and the magnetodisc. Evidence for the existence of this third major energetic particles from previous missions comes from the high flux levels observed by Pioneer 11 in the north high latitude Jovian magnetosphere (Kivelson, 1976; Kennel and Coroniti, 1979).

Various measurements collected from Ulysses throughout the dayside and, in particular, the duskside high latitude magnetosphere provided strong evidence about the existence of this large scale layer of energetic particles. Cowley et al. (1996) analyzed ~1 MeV proton fluxes from the Anisotropy Telescopes (ATs) instrument of the COSPIN experiment and first inferred that, beyond the inner magnetosphere, Ulysses sampled two large ion regions on its outbound pass of the Jovian magnetosphere: the well known magnetodisc plasma sheet in the middle magnetosphere and a large scale ion region in the outer magnetosphere, spreading from moderate latitudes to the boundary layer.

Later on, several researchers confirmed a significant contribution of flux gradients to the anisotropies in all the magnetospheric regions of Jupiter, by using measurements from different experiments onboard Ulysses. In the middle magnetosphere, the major gradient components were latitudinal and directed toward the equatorial plane, which is consistent with the flux maxima observed near/at the current sheet (Cowley et al. 1996; Staines et al. 1996; Laxton et al. 1997; Hawkins et al., 1998). On the contrary, the ion observations in the outer high latitude magnetosphere outbound revealed intensity gradients with positive latitudinal components, suggesting flux increasing in the direction away from equator, toward the southern pole (Laxton et al. 1997; Hawkins et al., 1998). These observations were noted as surprising results (Laxton et al. 1997).

Krupp et al. (1999) having analyzed in detail Ulysses/EPAC energetic ion measurements in the duskside high latitude magnetosphere outbound, reported a ~5 hr modulation in fluxes of ~0.57 MeV protons and ~0.59-0.66 MeV/N helium ions as well, and in the spectral slope of energetic protons. In order to explain the presence of ~5 hr variations, Krupp et al. suggested the existence of a boundary layer of energetic particles, which Ulysses encounters periodically, twice per rotation (~10 hours), due to the boundary's in-out oscillation.

In addition, Anagnostopoulos et al. (1998, 2001) presented an analysis of long lasting (~2-3hr) energetic ($\geq$~60 keV) ion and ($\geq$~40 keV) electron events observed by the Ulysses/HI-SCALE instrument in the high latitude Jovian magnetosphere, and inferred that these events (in both the prenoon dayside and the southern duskside magnetosphere) are different from the magnetodisc ion events. The HI-SCALE intensity, composition, spectral and pitch angle distribution measurements data were found to be, in general, consistent with the existence of large-scale layers of energetic ions and electrons in the high latitude Jovian magnetosphere (Anagnostopoulos et al., 1998, 2001, 2002; Karanikola et al., 2004; Marhavilas et al. 2004). In summary, the following observations were found to support a large scale boundary layer at high latitudes: (1) oppositely directed intensity gradients suggesting an additional ion layer northward or southward of the well known region of the magnetodisc, (2) an intensity minimum between low and high latitudes also suggesting another population at high latitudes, (3) spectral index increases as long as the spacecraft approached the magnetodisc and the magnetopause every ~10 hours, (4) different forms (in energy dispersion) of the flux-time profiles at high latitudes from those observed during the magnetodisc crossings, (5) different spectral characteristics in two adjacent particle regions (magnetodisc particle versus high latitude population), (6) a phase shift (of ~7 hr) between the spectral index peaks observed inbound and outbound, consistent with detection of a permanent layer northward/southward of the magnetodisc throughout the whole crossing time period of the Jovian magnetosphere that lasted from day 22 to 34, 1992 (Cowley et al., 1996; Hawkins et al., 1998; Laxton et al., 1997; Anagnostopoulos et al., 1998, 2001; Nishida, 2004; Woch et al., 2004; Krupp, 2004; Karanikola et al., 2004; Marhavilas et al. 2004).

The high latitude outer magnetosphere layer has been considered as the possible most important source of energetic particles escaping in the magnetosheath and therein, in the interplanetary space as far as ~1AU southward and northward of the Jovian magnetosphere [Zhang et al., 1993, 1995; Marhavilas et al., 2001; Anagnostopoulos et al., 2001, 2009].

Taking into account that this new layer plays an important role in the structure and the dynamics of Jupiter's magnetosphere, it is very important to further test and confirm its existence and structure. To this end, in this study we elaborate the values of the first order cross-field anisotropy of energetic ions in order to investigate the variable direction and the values of the cross-field intensity-gradients with regard to the instant position of the spacecraft. In particular, we examine the Ulysses' observations of high energy heavy ions (E>2.5 MeV, Z>5) and protons (0.5-0.95 MeV), which have large gyro-radius and allow a good comparison of the local flux variations with the remote sensing flux differentiation detected at the same time at the position of the spacecraft. We think that our present analysis provides very strong evidence that confirms previously presented evidence about the existence of a third major energetic particle (ion) region that was extended from middle to high latitudes (near the magnetopause), in both the dayside north (inbound) and the duskside south Jovian magnetosphere (outbound) during Ulysses' flyby of Jupiter in 1992.

## 2. Instrumentation

Ulysses spacecraft, a joint NASA and European Space Agency mission, was launched in October 6, 1990. To reach high solar latitudes, the spacecraft was aimed close to Jupiter so that Jupiter's large gravitational field would accelerate Ulysses out of the ecliptic plane to high latitudes. The Encounter with Jupiter occurred on February 8, 1992, and since then Ulysses travelled to higher heliospheric latitudes investigating for the first time the 3D Heliosphere. Although the primary mission of the Ulysses spacecraft was to characterize the heliosphere as a function of solar latitude, its flyby of the Jupiter provided the chance for investigating the duskside south Jovian magnetosphere for first time, or the high latitude magnetosphere for second time (after Pioneer 11). Ulysses officially ceased operations on June 2009, after receiving commands from ground controllers to do so.

Teams from universities and research institutes in Europe and the United States provided 10 instruments on board Ulysses. The Heliosphere Instrument for Spectra, Composition, and Anisotropy at Low Energies (HI-SCALE) onboard the ULYSSES spacecraft (Lanzerotti et al., 1992a) was constructed to investigate low energy ions and electrons. HI-SCALE consists of five apertures in two telescope assemblies mounted in a unit that contains the instrument electronics. To attain the lowest energy of response over a wide variety of particle species with appropriate geometrical factors and angular resolution, HI-SCALE utilizes three distinct silicon solid-state detector systems. These are Low-Energy Magnetic/Foil Spectrometers (LEMS/LEFS) and Composition Aperture (CA) [CA; is sometimes called "WART"]. The LEMS/LEFS systems provide pulse-height-analysed single-detector measurements with active anticoincidence. The CA system uses a multiparameter detection technique to provide measurements of ion composition in an energy range similar with LEMS/LEFS.

Energetic ions ($E_i>50$ KeV) and electrons ($E_e>30$ KeV) are detected by five separate solid-state detector telescopes, oriented to give essentially complete pitch-angle coverage from the spinning spacecraft. Ion elemental abundances are determined by a $\varDelta E$ vs $E$ telescope using a thin (5 μm) front detector element in a three-element telescope. Experiment operation is controlled by a microprocessor-based data system. In-flight calibration is provided by radioactive sources mounted on closable telescope covers. Ion and electron spectral information is determined using both broad-energy-range rate channels and a pulse-height analyzer for more detailed spectra.

Ulysses rotates around an axis which points toward Earth with a period of 12 sec. In the present study we use measurements of energetic proton channels W1 (0.5-0.95 MeV) and W2 (0.95-1.6 MeV) and high energy heavy ion channel Z3 (E>2.5 Z>5). The CA60 telescope detects particles in a direction that deviates $60^0$ from the spin axis of the spacecraft and collect measurements from 8 different sectors (each sector covers an angle of $45^0$).

## 3. Methodology

During Ulysses' flyby of Jupiter, the relative position of the sectors of telescope CA60, remained unchanged, with respect to a coordinate system positioned on the Sun. Careful analysis of CA60 / HI-SCALE observations within the Jovian magnetosphere suggests that the oppositely looking sectors counting large gyroradius ions positioned northward and southward from the position of the satellite are mainly sectors 4 and 8, as shown (Figure 1). This happens during both the inbound and outbound trajectory of the spacecraft. However, an anisotropy in the direction perpendicular to the magnetic field (first order anisotropy), for instance detection of much higher fluxes from sector 4 or 8, can be detected due to two main reasons: either to the detection of energetic particle intensity gradients within a region with spatial variation in particle intensities perpendicular-to- the magnetic field, or to the transportation of an energetic particle population due to the magneto-plasma motion in the direction of the highest detected intensities, which is known as the Compton-Getting effect (Northop, 1979). In this study we use the evaluation of the cross-field ion anisotropy vector in order to elaborate the intensity gradients and test the existence of the two separate particle regions (magnetodisc - high latitude particle layer) northward and southward of Ulysses, during both the inbound and outbound segments of its trajectory in February 1992. We will show that in the cases examined, the Compton-Getting effect made negligible or very small contribution.

In order to better explain the use of the cross-field ion anisotropy vector as a tool for monitoring the energetic particle intensity gradients within a region with spatial variation in particle intensities perpendicular-to-the magnetic field, we present in Figure 2 a sketch indicating the way of study of the dayside large scale high latitude energetic particle layer during the inbound trajectory of Ulysses. More explicitly, Figure 2 shows a schematic representation of the two separate energetic particle populations, the magnetodisc-ion layer and the high-latitude ion layer (Cowley et al.,1996; Anagnostopoulos et al., 1998, 2001; Marhavilas et al., 2004) in the XY plane, with respect to the position of the spacecraft detector (presented by symbol " )( "). The X and Z axes in this scheme are located on the ecliptic plane, with Z axis pointing toward the Earth (and the Sun), X toward the dawn Jovian magnetosphere, and the Y axis completes the coordinate system. Ulysses traveled at moderate latitudes on the line connecting Jupiter with Sun while the magnetic field in the north hemisphere of the outer magnetosphere was almost parallel to the ecliptic plane, in the general direction away from the planet, near the X axis direction. From Figure 2 we can see that for a position of the spacecraft at the edge of the outer high latitude magnetosphere ion layer, more ions enter the CA60/HI-SCALE head from sector 8, which detect ions gyrating northward of its position, whereas for a position of the spacecraft at the north edge of the magnetodisc, more ions enter the CA60/HI-SCALE head from sector 4, which detect ions gyrating southward of the spacecraft trajectory inbound. When the spacecraft crosses the magnetodisc and is found in its south edge, more ions will again enter in the direction of sector 8. We infer that the magnetic field and the looking directions of sectors 4 and 8 are such as that when an anisotropy between sectors 4 and 8 (counting particles perpendicular to the magnetic field) is observed, then an intensity gradient from south-to-north or north-to-south direction can be present.

In next paragraphs we will present and discuss Ulysses measurements made during its inbound trajectory in the north dayside magnetosphere (Figures 3 and 4). If our scenario for the periodic Ulysses' approach to two separate regions is correct, then an increase in the intensity of sector 4 that results to a positive value of the anisotropy index

$$A = \frac{S4 - S8}{S4 + S8} \quad (1)$$

will suggest an intensity gradient from north to south, that is an approach to the plasma sheet. On the contrary, a relative increase in the intensity of sector 8 will result to a negative value of the anisotropy index A that will be consistent with an intensity gradient in the direction from south to north, toward the high latitude particle layer. More details are discussed in Section 4.

As we explained, according to Figure 2, for the north hemisphere of Jupiter, the positive values of index A indicate the approach to the plasma sheet, while the negative values of A suggest the existence of ion intensity gradients toward the north, suggesting the existence of a particle concentration at higher latitudes. Assuming absence of magneto-plasma transportation across the magnetodisc (Compton-Getting effect) in the X axis direction and a linear pattern at the sites of flux variations above the dayside plasma sheet, the relative intensity gradient can be written as follows:

$$RG = \frac{|\vec{\nabla} j|}{j} \cong \frac{\frac{j_{PS} - j_{OL}}{2R_g}}{\frac{j_{PS} + j_{OL}}{2}} = \frac{1}{R_g}\left(\frac{j_{PS} - j_{OL}}{j_{PS} + j_{OL}}\right) = \frac{1}{R_g}\left(\frac{S4 - S8}{S4 + S8}\right) = \frac{A}{R_g}$$

(2)

where $j_{PS}$ and $j_{OL}$ are the intensities of the particles (in the neighborhood) of the plasmasheet and at north higher latitudes, respectively, S4 and S8 are the particle intensities measured by sectors 4 and 8, $R_g$ is the gyroradius of the particles and *A* is the anisotropy index. The equation (2) suggests that as the bigger the cross-field anisotropy A is as the stronger the relative intensity gradient $|\vec{\nabla} j|/j = A/R_g$ becomes. Assuming that Ulysses counted $O^+$ ions, we get

$$R_g = \frac{mv_\perp}{qB} \cong 0.88 R_J \quad (3)$$

where $R_J$ is the Jovian radius, $v_\perp$ is the particle velocity and m, q the mass and the electric charge of the particle.

It worth to note that both segments of Ulysses trajectory during its flyby of Jupiter in 1992, that is the inbound and the outbound ones, provided complementary chances for the elaboration of the high latitude outer magnetosphere large scale layers. The Ulysses' inbound trajectory was proved to be very useful for our analysis based on the study of the intensity gradients perpendicular to the magnetic field, because during this period the spacecraft traveled between the magnetodisc and the high-latitude particle region and a change in the direction of the intensity gradient vector due to its ~10hr periodic approach to each region constitutes a crucial test for the investigation of the energetic ion spatial distribution. The ~10hr periodic approach of the spacecraft to the magnetodisc has been easily determined by the increase of the energetic particle intensity and the decrease in the magnetic field magnitude. Our analysis extends our knowledge of the structure of the Jovian magnetosphere by investigating an energetic ion layer northward of the plasma sheet. In addition, during the time period that Ulysses crossed the particle region of high intensities in the south high latitude magnetosphere outbound (Cowley et al., 1996; Anagnostopoulos et al., 2001), the information gained from the measurements made therein, provides the possibility of double checking the particle spatial distribution: by direct detection of the flux variation and indirect estimation of the intensity gradients.

In the data analysis to be presented in the next section we use appropriate HI-SCALE heads and energy channels, which detect ions with large gyroradius and allow a good remote sensing of the cross-field intensity gradients.

## 4. Measurements of ion intensity gradients gradients in the High Latitude outer magnetosphere

**4.1 Distinction between energetic ions in the Magnetodisc and the High Latitude Layer in the middle magnetosphere: Inbound high energy heavy ion observations**

We start our data analysis by examining the cross-field anisotropy and the corresponding intensity gradients at the times of Ulysses plasma sheet crossings, where the ion behavior has been well described in the relative scientific literature (Van Allen et al.,1974; Simpson et al., 1974; Kivelson, 1976; Kennel and Coroniti, 1979; Krimigis et al., 1981; Zwickl et al., 1981; Lanzerotti et al.,1993). First we focus on day 37 of year 1992, because of some ~10 hour quasi-periodic large flux increases of energetic particles associated with magnetic field magnitude decreases and field direction $180^0$ changes demonstrating Ulysses plasma sheet approaches and subsequent current sheet crossings.

In the top panel of Figure 3, we show the flux-time profile of heavy ions as measured from channel Z3 (Z>5, E>2.5 MeV) of the CA60/HI-SCALE detector. The second panel presents the intensities S4 (black line) and S8 (grey or red line in the coloured version of the paper) measured by the two oppositely looking sectors 4 and 8, which, as we mentioned in the previous section, count particles perpendicular to the magnetic field. Panel (c) illustrates the anisotropy index A (= (S4-S8)/(S4+S8)), which, in particular at times of abrupt flux variations near the plasma sheet, we consider to indicates the particle cross-field energetic ion anisotropy. In panel (d) we display the relative intensity-gradient RG=$|\vec{\nabla}j|/j$ evaluated by the index A according to the equation (2); in panel (d) we also present one hour average values of the relative intensity gradient (grey or red line). The two bottom panels (e) and (f) of Figure 3 illustrate the magnetic field magnitude B and the angle $\phi$ in a spherical coordinate system.

Figure 3 suggests that throughout day 37, three large ~10 hr periodic flux enhancements of heavy ions were observed: one between ~0100-0400 UT, a second between ~0730 – 1600 UT and a third after ~2000 UT. During the two first ~10 hour separated intervals, with increases of the particles intensity, decreases of the magnetic field magnitude B and reversal of its direction from sunward to anti-sunward, were observed. These two ion events are well identified as ion flux increases due to current sheet crossings (Cowley et al., 1996; Krupp et al., 1993); Anagnostopoulos et al. 1998, 2001). The third increase in the particles intensity (observed after 2000 UT), is not accompanied by a reversal in the direction of the magnetic field (the angle $\phi$ remained at almost the same values), suggesting that in this case, Ulysses approached but did not cross the current sheet.

What is most important in Figure 3 for the present study is the two ~10 hour separated flux increases observed between the successive plasma sheet events, which are inconsistent with flux minima anticipated by the "magnetosdisc model". More explicitly, the reader can see in the top panel of Figure 3 that the weaker plasma sheet ion events were centered at ~0530 UT and ~1730 UT. The intensities at these times are lower than the plasma sheet events, but they are much higher than the ones measured in both the interplanetary space and the magnetosheath (Anagnostopoulos et al., 1998, 2001; data not shown here). Furthermore, these weaker events observed between the times of the plasma sheet flux enhancements were found to be associated with distinct energy spectral index peaks and different characteristics than the plasma sheet ion events (Anagnostopoulos et al., 1998, 2001). These heavy-ions intensity increases, detected by Ulysses outside the dayside magnetodisc, provide a first evidence for the existence of an energetic ion population at higher latitudes than the position of the magnetodisc.

We have already mentioned that the large increase in the intensity of heavy ions at the beginning (0000-0130 UT) of day 37 shows the Ulysses approach and subsequent crossing of the magnetodisc. We also note that after Ulysses traversed the plasma sheet, it remained from ~0145

UT to ~0300 UT, in the south hemisphere, as suggested from the magnetic field measurements (panels e and f). The B field magnitude and direction suggest that after ~0300 UT Ulysses returned to the north hemisphere; after this moment, from ~0300 UT to ~0430 UT, the intensity measured from channels Z3 and the sectors 4 and 8 gradually decreased, suggesting a removal of Ulysses from the plasma sheet toward higher latitudes at this time interval. At ~0430 UT (dash line A) Ulysses observed a short Z3 flux minimum (Panel a), but a careful examination of the intensities of sectors 4 and 8 (Panel b) reveals that the S8-intensity is much less than S4-intensity. The relative decrease of S8 comparatively to S4 at the end of the plasma sheet ion event, results a positive value in the anisotropy index A seen in Panel c suggesting an intensity gradient vector looking toward the plasma sheet; the relative intensity gradient (panel d) at that time was quiet large: ~40% $R_J^{-1}$. At that time, when the anisotropy-index peak and the strongest southward intensity gradient were detected by Ulysses, the spacecraft seemed to have left the influence of the magnetodisc.

Concerning the main scope of this study, it is important to note the values of S4, S8, A and RG after ~0430 UT up to the time of influence of the next plasma-sheet approach at ~0730. In particular, the situation at ~0545 UT is exactly opposite than that at ~0430 UT. First, the whole flux structure from the Z3 measurements (panel a) coincides with the flux structure shown for the S8 ions, which has been detected by the Ulysses CA60/HI-SCALE head from the region northward of the spacecraft satellite. Moreover, the maximum increase in the Z3 and S8 high energy heavy-ion intensity at ~~0545 UT coincides with a minimum in S4 intensity of ions detected by the detector from south (solid circles in S4 and S8 fluxes at a time indicated by the dashed normal line marked with B. This form of Z3, Z3-S4 and Z3-S8 intensity flux-time profiles strongly suggests that the population found at high latitudes controls the Z3 ion-observations from (Cowley et al., 1996; Anagnostopoulos et al. 1998, 2001; Marhavilas et al., 2004). As a consequence, at ~0545 UT, the evaluated anisotropy index and the relative intensity gradient alter their sign relatively to that seen at the edge of the plasmasheet at ~0430 UT, and show negative values, with an intensity gradient toward the north as much as ~90% $R_J^{-1}$, an almost doubling in ion flux increase within a Jovian radii. Later on, a minimum Z3 flux at ~0730 UT (dashed line Γ) was followed by an increase lasting for more than 2 hours as Ulysses approached the current sheet. At the time of Z3 flux minimum at ~0730 UT, a negative minimum value of the anisotropy index A (panel c) and of the relative intensity gradient RG (panel d) was observed, followed by values near zero; these measurements are consistent with a removal of Ulysses from the high latitude ion region (strong northward intensity gradient) and a subsequent approach toward the magnetodisc. Actually, the indicated in panels e and f, magnetic field measurements suggest a first approach to the plasma sheet at ~1000 UT and current sheet crossings at ~1130 and ~1200 UT.

A similar situation with a distinct region of energetic ions observed at north high latitudes between the successive ~10 hour separated plasma sheet crossings centered at ~0230 UT and ~1130-1200 UT can be seen during the next removal of the spacecraft from the magnetosdisc. The variation of the Z3 energetic-ion high flux levels (panel a) along with the magnetic field directional reversals and field magnitude decreases (panels e and f) are signatures of the magnetodisc control on the Z3 ions up to ~1500 UT. Furthermore, around ~1500 UT the sector-4 of CA60 was detecting more particles (coming from the plasma sheet) than CA60 sector-8 ions (reaching from north) under a quiet magnetic field of increasing magnitude. These observations imply a strong positive value of the anisotropy index A and southward intensity gradient, which suggest that most probably Ulysses at ~1515 UT crossed the edge of the main magnetodisc ion concentration (peaks in panel c and d marked by solid lines at the time indicated by the dashed normal line Δ). The important measurements for the scope of the present paper are those during the time interval ~1615-~2000 UT. The Ulysses CA60/HI-SCALE head observed during this time interval in general higher S8 than S4 fluxes, and negative anisotropy and intensity gradients suggesting entrance of Ulysses for about 4 hours in an ion region northward of the plasma sheet, at higher latitudes, as in the previous exit from the plasma sheet (centered at ~0600 UT). In particular we note the peak at negative values of the anisotropy index A and the intensity gradient RG (strong northward gradient vector direction) at ~1730 UT (dashed line E) by the increase of S8 flux and the decrease of S4 flux (panel b), which have been simultaneously observed with an increase of the total Z3 flux (panel a). These observations strongly suggest that Ulysses approached the south edge of the north high latitude ion region reported in previous studies (Cowley et al., 1996; Anagnostopoulos et al. 1998, 2001; Marhavilas et al., 2004); a similar approach seems that was occurred at ~1915 UT (normal dashed line Z). The spacecraft removal from the high latitude layer and its following approach to the magnetodisc, a Z3 ion population is strongly suggested by the abrupt Z3 flux increase at ~2000UT, associated with the positive A and RG maxima at the same time; notice that the evaluated relative intensity gradients were ~60-70% $R_J^{-1}$ for all the four A and RG peaks marked with Δ, E, Z and H.

Figure 4 is constructed in the same format with Figure 3, for the time interval 15:00 – 24:00 UT, day 36, 1992. According to the analysis and he discussion of Figure 3, the peak A (panel c) in the values of A and RG at ~2010 UT indicates the time that, when Ulysses moved away from the magnetodisc, controlled Z3 ion population. Most important, the negative values of the anisotropy index A and of the relative intensity gradient RG between ~2100 – 2400 UT (in particular, Peaks

B, Γ and Δ, in panel c) confirm the detection of the high latitude layer of energetic particles during the period of that exit from the magnetodisc on day 36, 1992 as well.

**4.2 Distinction between energetic protons in the Magnetodisc and the High Latitude Layer in the middle magnetosphere outbound**

Now we examine the northward-southward intensity gradients by studying the cross-field ion anisotropy in the middle Jovian magnetosphere on Ulysses outbound trajectory. Because of the rapid southward direction of Ulysses trajectory outbound, repeated current sheet crossings were not detected. Therefore, as the "middle magnetosphere" has been considered, the regime traversed by Ulysses from the beginning of day 40 until ~2000 UT on day 41, ~50 R$_J$, is characterized by 10-hour modulations of the ion flux, with the highest fluxes observed closest to the current sheet at low latitudes, associated with depressions in the magnetic field strength (Balogh et al., 1992; Lanzerotti et al., 1992b, 1993; Philipps et al., 1993; Cowley et al., 1996; Anagnostopoulos et al., 1998, 2001). Since the Z3 high energy heavy ion fluxes were low and the flux increases not clearly distinguishable between the middle of day 40 and of day 41 (Lanzerotti et al.,1993; their Figure 6), we proceed to examine the northward-southward intensity gradients by studying from the HI-SCALE/CA60 W1 high energy proton observations.

As in the case of CA60 Z3 heavy ions, the oppositely looking sectors of measuring W1/CA60 proton fluxes in a direction perpendicularly to the magnetic field are the sectors 4 and 8 (Figure 1). This is due to the fact that the arrangement of the CA60 sectors in the duskside magnetosphere of Jupiter was remained unchanged relatively to the inbound trajectory. Similarly the magnetic field direction, during both the inbound (north dayside) and outbound (south duskside) trajectories was almost in the same direction (sunward). On the contrary the locations of the plasma sheet and the high latitude particle layer relatively to the position of the spacecraft were reversed during the outbound trajectory, with the particle layer northward of Ulysses and the high latitude layer southward of the spacecraft.

Figure 5 displays the intensity profile of W1 protons, in panel a, the intensities of the two oppositely looking sectors 4 and 8 (black and grey line respectively), in panel b, the cross-B field anisotropy index A, in panel c, and the relative intensity gradient RG (the red line corresponds to 1hr-averaged values), in panel d, for the time interval 0000, d40 - 0600, d41/1992. The channel W1 records protons in the energy range 0.5-0.95 MeV, and by considering the geometric mean energy, we use in Eq. 1 a proton energy value of E=0.68 MeV and the proton gyroradius is evaluated to be R≈0.17R$_J$. Following to the previous discussion on the measurements of W1/CA60 channel within the south duskside magnetosphere, we note that a higher intensity in sector 4 than in sector 8 reflects a southward intensity gradient (high latitude particle layer), while the opposite situation suggests a northward intensity gradient (plasma sheet).

In Figure 5 the large increases in the intensity at 0300-1000 UT /d40, 1300-2100 UT /d40, and 0000-0600 UT /d41 are caused by the repeated approaches of Ulysses to the plasma sheet region. For these intervals a negative value in the anisotropy index was detected, which indicates a gradient in a direction toward the plasma sheet (northward). In particular, we note as an example, that the calculated relative intensity gradient at ~2200 UT, d40 was found to be as high as ~160 % $R_J^{-1}$.

Between the W1 proton events centered at ~0800/d40, ~18UT/d40 and 0200UT/d41 (Figure 5), Ulysses observed some fluctuations, but at flux levels much higher compared to those observed in the magnetosheath and upstream from the bow shock (Figure 5a, 5b). At those times Ulysses moved southward (see below in the text related with Figure 7) and approached the high latitude particle layer in the outer magnetosphere (Anagnostopoulos et al., 2001). Indeed, for instance, between ~1110-1230 UT and ~2100-2400UT on day 40/1992, positive values of the anisotropy ratio A and the intensity gradient RG are seen, which are consistent with the existence of a W1 proton population southward of the spacecraft. In the next section we examine in detail this ion population for the whole time period while Ulysses crossed a large scale layer at south high latitudes from near the end of day 41 (~2000 UT) until the first outbound magnetopause crossing in the second half of day 43 at ~83 $R_J$ (Cowley *et al*. 1996; Anagnostopoulos et al., 2001), in the outer magnetosphere.

## 5. Further observational testing of ion intensity gradients in the High Latitude outer magnetosphere

In this section we examine Ulysses high energy ions from the HI-SCALE instrument and relativistic electron measured by the COSPIN experiment, during the segment of the spacecraft trajectory in the Jovian magnetosphere outbound, as long as it crossed the characteristic particle structure at high intensity, identified as the outer magnetosphere. The new thing of this section, beside the detailed study of the cross-field anisotropy probably related with southward intensity gradients, is that we perform four different tests to confirm the hypothesis of the principal contribution of intensity gradient to the anisotropies under examination.

**Test 1: Comparison of flux and cross-field anisotropy direction changes**

In the first panel (a) of Figure 6 we present the Z3 high energy heavy (Z>5, E>2.5 MeV) ion intensities obtained by the Ulysses CA60/HI-SCALE telescope for the time interval 1500 UT, d41 – 1200 UT, d43 of 1992. In the second panel (b) we show the intensities measured by sectors 4 and 8, while in the third panel we have evaluated the anisotropy index A (Section 3, Eq. 1). We remind that we have explained above (discussion on Figure 5) that sector 8 records large angle ions arriving from regions located northward of Ulysses, while sector 4 records ions from regions located southward of its instant position. From Figure 6 we see that the anisotropy index A shows strong variations between positive and negative values when Z3 flux shows minimum values at ~1600UT/d42, ~0100UT/d43, ~1030UT/d43 (normal dashed lines in Figure 6). These three events are separated by a time interval of 9.0-9.5 hours, which is near the ~10 hour rotation period of Jupiter and they have been identified as times of Ulysses approaches to (or crossings traverses of) the magnetopause (Phillips et al., 1993)). Since the ~10 hour periodic variation in the anisotropy index A (panel b) is related with both Ulysses' approach to the magnetopause and the Z3 ion flux minima (panel a), we infer that the cross-field anisotropy is well related with ion intensity changes obviously due to strong intensity variations near the magnetopause and cannot be attributed to the Compton-Getting effect. Furthermore, a comparison of the general changes in all the time profiles of Z3 (panel a), S4 and S8 fluxes as well as the cross-field anisotropy A, for the whole period of the enhanced ion flux in the outer magnetosphere from ~2100 UT, d41 until 1000 UT, d43 of 1992, also suggest a good relation, which implies that the cross field anisotropy is mostly due to intensity gradients, evidently caused by the ~10 hour quasi-periodic motion of the Jovian magnetosphere.

**Test 2: Comparison of ion versus electron cross-field anisotropies.**

We further examine the nature of the cross-field anisotropy, and so that we perform a crucial test on Ulysses observations in the south outer magnetosphere outbound by focusing on a comparison of ion and electron anisotropies, as shown in Figure 7, for the time interval 1200 UT, d42 – 0300 UT, d43. The three panels of Figure 7a display the same quantities as in Figure 6, while the three bottom panels of Figure 7b illustrate the similar quantities with Figure 7a but for relativistic (E>3 MeV) electrons measured by the H7S channel of the COSPIN experiment onboard Ulysses. The two vertical lines in Figure 7a correspond to the first two vertical lines of Figure 6, and indicate times when Ulysses was at the magnetopause, where a decrease in the intensity of energetic particles was observed. In the same figure we display the intensity profiles of sectors 4 (black line) and 8 (grey line) of the detectors CA60/HI-SCALE and HET/COSPIN as well.

As in the case of heavy ions, whenever Ulysses was approaching the magnetopause, a decrease in the intensity of electrons is observed as well as in the intensities of sectors 4 and 8, with higher decrease in the intensity of sector 4 that causes the 10hr periodic variation of the anisotropy. If we take in mind that the electrons gyrate oppositely to ions, and the fact that sectors 4 and 8 of CA60 and HET detectors are located exactly oppositely, the anticipated anisotropies should be of the same sign in the case of detection of intensity gradients, but of the opposite sign due to Compton Getting effect. We clearly see that the cross-field Z3 ion anisotropy decrease at the first magnetopause approach (~1530-1620 UT, d42) presents the same sign (negative) with that of the HET relativistic electron anisotropy. A similar behavior is seen during the second flux decrease (~0015-0100 UT, d43) at the next magnetopause approach (except for the instant of a small local maximum in the HET electron flux within the second intensity dip, when the anisotropy vanishes).

In conclusion, the crucial test of comparing the ion anisotropy sign with the electron one verifies that the observed anisotropy in these cases are caused by the intensity gradient and not by the Compton-Getting effect.

**Test 3: A comparison of small versus large energy ion anisotropy.**

A first order cross-B field anisotropy can be observed in energetic particle pitch angle distributions due to two factors: (a) a bulk flow of particles with respect to the spacecraft (Compton-Getting effect) or/and (b) the existence of a spatial intensity gradient. The first order anisotropy is given by the relation

$$\mathbf{A}_1 = R_g \hat{\mathbf{b}} \times \frac{\vec{\nabla} f}{f} + \frac{1}{f} \frac{\partial f}{\partial v} \mathbf{V}_c$$

(4)

(Northrop et al., 1979), or

$$\mathbf{A}_1 = \frac{R_g}{L} \hat{b} \times \hat{n} + \frac{2(\gamma+1)}{v} \mathbf{V_c}$$

(5)

if a power law function (f ~ $E^{-\gamma}$) of the spectrum is considered (Carbary et al. 1981; Kane et al.,1995), where $\mathbf{A}_1$ is the first-order anisotropy vector, $R_g$ is the particle gyroradius, $\hat{b}$ is the unit vector of the magnetic field, f is the measured particle flux, v is the particle speed, $\mathbf{V_c}$ is the bulk flow velocity of particles with respect to the spacecraft, $\hat{n}$ is the unit vector in the direction of flux gradient, $L$ is the scale length of the spatial intensity gradient and γ is the spectral index.

According to the first term of equation (4) the ion anisotropy resulted from intensity gradient, is proportional to the particles gyroradius, and therefore, it is more sensitive for high energy heavy ions. According to the second term of equation (4), the anisotropy that is caused by the Compton-Getting effect, is inversely proportional to the particle speed (energy), that means it is more intense for particles of low speed (lower energy).

In Figure 8 we display representative pitch angle distributions of particle intensities from the HI-SCALE instrument onboard Ulysses, in two columns corresponding to measurements at two different times, as long as Ulysses traveled near the dusk-side south magnetopause outbound (days 42 and 43, 1992). Each column displays pitch angle distributions in the region from $0^0$ to $180^0$ for three kind of detected froms: the high energy heavy ion CA60 / Z3 (Z>5, E>2.5 MeV) channel (top), the low energy proton CA60 / W1 (0.5-0.95 MeV) channel (bottom), and the high energy proton CA60 / W2 (0.95-1.6 MeV) channel (middle). The numbers (from 1 to 8) in each frame indicates the sector from which the ion intensity (vertical axis) at the corresponding pitch angle (horizontal axis) was measured. We see that sectors 8 and 4, which are looking in opposite directions (Figure 1) are measuring ion intensities with pitch angle of ~$90^0$, that means particles gyrating perpendicular to the magnetic field direction, and that S8 flux is higher that S4.

In Table I we have evaluated the gyroradius and anisotropy (Eq. 1), at ~$90^0$, for the mean energy of the proton channels W1 and W2, and the high energy heavy ion channel Z3 (assumed Z=5 for the gyroradius calculation), in each of the six measurements displayed in Figure 8. It is evident that the anisotropy magnitude A is ordering according to the gyroradius value $R_g$, with increasing A magnitude as $R_g$ increases. This ordering is consistent, according to the previous analysis (Eq. 4) with cross-field anisotropy caused by ion intensity gradient, and not by the Compton-Getting effect.

-16-

**Test 4: Comparison of time varying particle intensities with instant measurements of intensity gradients (inferred from cross-field anisotropies).**

A final test on cross-field anisotropy as measuring the ion intensity gradient in specific time intervals of intensity changes throughout Ulysses' flyby of Jupiter is performed for the plasma sheet edges in the south magnetosphere outbound. During Ulysses outbound trajectory the spacecraft traveled very fast toward higher south latitudes in the duskside magnetosphere. The $r_x$ and $r_y$ components of the spacecraft position vector **r** did not change significantly, and we consider that the change of Ulysses position approximately equals the spacecraft removal along z axis, that is $\Delta r \approx \Delta r_z = \Delta z$, for a coordinate system positioned on the planet (where Z is the rotation axis of the planet, X is the axis lying on the ecliptic plane looking from Jupiter toward the Earth, and Y is the axis looking from the dawn to the dusk magnetosphere).

As a first example, we study below the case of Ulysses removal from the dusk plasma sheet on day 41. Assuming a linear intensity variation along the z axis, an approximate evaluation of the relative change of the intensity gradient results from the relation:

$$\frac{|\vec{\nabla}j|}{j} \cong \frac{\frac{j_{PS} - j_{OL}}{\Delta z}}{\frac{j_{PS} + j_{OL}}{2}} = \frac{2}{\Delta z}(\frac{j_{PS} - j_{OL}}{j_{PS} + j_{OL}}) = \frac{2A}{\Delta z} \tag{6}$$

where A is the anisotropy index. We apply the above relation for the gradual W1 (0.5-0.95 MeV) proton decrease, concerning the time interval 1700-2000 UT of day 40 (panel a of Figure 5), when the direction of the gradient vector is continuously negative (panel d of Figure 5), indicating more ion concentration northward of the spacecraft. By putting $j_{PS} \approx 140$ and $j_{OL} \approx 19$ (panel b of Figure 5), the average relative intensity gradient during the four hour interval 1700-2000 UT (panel d of Figure 5) gives

$$RG = \frac{|\vec{\nabla}j|}{j} \approx \frac{2 \cdot 0.76}{2R_J} \approx 0.76 R_J^{-1} \ .$$

Then we evaluate the value of RG from the intensity difference and Ulysses removal Δz along the z axis (panel a of figure 5) and we take RG ≈ 0.77 $R_J^{-1}$, that is quite the same with the value found above.

We applied the same method in the case of a smoother flux variation (decrease) within the high latitude region of energetic particles in the south dusk-side outer magnetosphere, for the time interval 1200-1300 UT on day 42 (see Figure 9 in the next section). By applying the relation (6), the average relative intensity gradient for the time interval of interest was calculated to be RG ≈ 0.44 $R_J^{-1}$, while from the flux variation for the corresponding Δz Ulysses removal turns up to be 0.46 $R_J^{-1}$.

The excellent agreement of the values of the relative intensity-gradient derived by performing two different calculations, that is one by using relation (6) giving the remote sensing intensity gradient, and by direct deriving of RG from the flux variation for the corresponding Δz Ulysses removal strongly suggest that the observed cross-B field anisotropy in these two cases, at the edge of the plasma sheet and within the outer magnetosphere during the spacecraft outbound trip in the Jovian magnetosphere was mainly caused by ion intensity gradients and not by the Compton-Getting effect.

## 6. The magnetopause boundary layer of energetic particles

In this section we concentrate our attention on Ulysses observations of W1 energetic protons around the magnetopause as the spacecraft moved to exit the duskside south Jovian magnetosphere. Figure 9 has been constructed in the same format of Figure 5, but for the time interval 0900-2400 UT of day 42, 1992; on day 42 Ulysses traveled the region with the enhanced flux level, which has been identified as the outer magnetosphere or the high latitude particle layer. However, although the particle intensity achieved high values, they did not remain at a constant level, neither displayed a monotonic variation. On the contrary, as the relative literature has noted, the flux time profile is variable mostly due to the ~10 hour periodic motion of the magnetosphere (Marvavilas et al., 2004). For instance, the flux minimum seen in Figure 5 at ~1610 UT (~1600 UT) was detected at a time when Ulysses, in the course of its ~10 periodic motion in the south magnetosphere, moved to highest south latitudes and observed a local peak of the spectral index peak (Marhavilas et al, 2004; their Figure 5); this happened quasi-periodically every ~10 hours as long as Ulysses moved southward to leave the magnetosphere (Marhavilas et al., 2004; their Figure 5). Therefore, the ~10 hour quasi-periodic southward motion of Ulysses suggests that the spacecraft approached quasi-periodically the south boundary of the magnetosphere: the magnetopause. We understand the flux maximum centered at ~1200 UT as observed close to the actual intensity maximum across the high latitude ion layer. After that time the proton intensity decreased almost continuously for about three hours (panel a), while relatively higher intensity was measured in sector 8 than in sector 4 (panel b) and negative values of the anisotropy index were observed (panel c) indicating an intensity gradient in the northward direction. This set of data is self-consistent with a maximum at moderate latitudes and flux decrease toward the magnetopause boundary. What is new in Figure 9 is the flux maxima centered at ~1500UT and ~1615UT.

We see that after ~1515UT the Ulysses CA60 / W1 proton flux decreases suddenly and is related with a peak in positive values of the anisotropy index A (peak A in panel c) indicating a strong southward intensity gradient, in the general direction of magnetopause. Then, an intensity minimum was detected (panels a and b) followed by a reversal of the gradient vector (panel d; marked as B in panel c) suggesting that the spacecraft was moved southward of a particle layer. After that time, Ulysses observed again a second short flux maximum, at ~1615 UT (panel a) accompanied with a peak in positive values of the anisotropy index A (peak Γ in panel c) indicating a recovery of the particle layer below the spacecraft (positive-southward intensity gradients in panel d). We identify this layer that crossed by Ulysses at that time as the magnetopause energetic particle layer. A similar effect was detected during the time interval ~2100-2300 UT. Then Ulysses approached again (peak Δ) and crossed (peak E) the energetic particle magnetopause boundary (Phillips et al., 1993), based on plasma measurements, identified the Ulysses detection of the magnetopause boundary layer a bit later, at 0024UT/d43.

A schematic representation of the W1 (0.5-0.95) MeV proton layers that Ulysses observed during its outbound trajectory, as long as it moved toward south higher latitudes to leave the Jovian magnetosphere and start its long journey in the 3D Heliosphere, is shown in Figure 10. In this sketch we suggest that Ulysses crossed three different particle regions: the well known plasma sheet (magnetosdisk), the high latitude layer (outer magnetosphere) and the magnetopause boundary layer.

## 7. Conclusions

In this study we examined the intensity gradients of heavy ions (Z>5, E>2.5 MeV) and protons (0.5-0.95 MeV) during the entire encounter of Ulysses with the Jovian magnetosphere in February of 1992. Our analysis verifies the existence of a particle layer at high latitudes during both the inbound and outbound Ulysses' trajectories, by studying the first order anisotropy of the energetic particles' intensity-gradient, perpendicular to the magnetic field.

In Figure 11 we show a model of the Jovian magnetic field (Connerney et al., 1981) and the trajectory of Ulysses in magnetic coordinates, during its flyby of the planet. The Z axis denotes the distance along the magnetic dipole axis, while the X axis shows the distance perpendicular to the Z axis. The big circles on the trajectory curve indicate the beginning of the day, while the dots denote the 2hr time-intervals. The arrows in the figure have been positioned on the trajectory curve at times of Ulysses removal from the magnetodisc toward higher latitude as long as the spacecraft moved in the middle magnetosphere, both inbound and outbound. These arrows at high latitudes, also indicate times (around 2200 UT/d36, 0545 UT/d37, 1730 UT/d37, 1145 UT/d40, 2200 UT/d40 and 1800 UT/d41) for which our analysis in this paper demonstrated local flux increases and simultaneous cross-field anisotropies strongly suggesting significant intensity gradient to the same direction of spacecraft motion (toward even higher latitudes). The permanent simultaneous measuring of flux increase and intensity gradients toward the high latitude outer magnetosphere suggests that the flux increases at these times imply that the intensity fluctuations (increases) observed in the middle magnetosphere should be attributed to a spatial and not to a temporal effect. The intensity gradients of the ion events observed between plasma sheet events indicate the inner ("near" the plasma sheet) edge of the large scale high latitude particle layer. Due to the geometry of Ulysses trajectory, the spacecraft had the chance of investigating the whole high latitude magnetosphere in its duskside south portion and crossing the high latitude energetic ion layer. Several authors in previously published papers have noted that the Ulysses particle intensities in the high latitude layer were at such high level as the intensities within the magnetodisc or even higher. We note that the arrows indicated in Figure 11, concern only the examples analyzed in the previous sections of this study, and they are representative of the effects found in Ulysses HI-SCALE measurements during the spacecraft flyby of Jupiter. We infer that the high latitude ion layer was a permanent feature of the Jovian magnetosphere for the whole time period that Ulysses spent in the middle and the outer magnetosphere.

However, we should address the following questions. First question: is it possible that the particle intensity minimum observed between the magnetodisc and the high latitude layer, is due to a particle absorption from Ganymede's magnetosphere? Second and more general question: is the high latitude particle layer a different region from the magnetodisc or a kind of continuation of the magnetodisc population? Concerning the first question, we note that Ganymede is placed on a distance of ~15 $R_J$ (14.97 $R_J$) from Jupiter, and Figure 11 shows that the magnetic field lines passing through the high latitude layer reach the ecliptic plane on distances much longer than the position of Ganymede. Therefore, the flux minimum between the magnetodisc and the high latitude layer is almost impossible to produce this large scale effect in the Jovian magnetosphere. Concerning the second more general question, we should take into account that a series of observational characteristics are absolutely different in the two regions (Anagnostopoulos et al., 2001). The ion events appear different relation between spectral index peaks and flux maxima, different energy dispersion in their flux-time profiles, etc. Furthermore, the high latitude layer is characterized by ~15-20 min and ~40 min periodicities and seems to have as its main source this kind of periodic emissions from near the planet at high latitudes. An estimation of the ion emission assuming energetic (50 keV) protons and in a cylindrical geometry (of the high latitude layer) between 50-80 $R_J$, from a measured density $n \approx 0.94 \times 10^{-4}$ particles/cm$^3$, implies a number rate of dN/dt =$10^{28}$ particles/sec, which is equal to the source rate of Io heavy ions, and consistent with the finding that the proton and heavy ion densities are almost equal (Krimigis et al., 1979).

We should notice that a sufficient number of investigators have studied the first order anisotropy perpendicularly to the magnetic field, during the Ulysses' flyby of Jupiter, and the results of their study seem to support the existence of an energetic particle layer at high latitudes (Staines et al. 1996; Laxton et al. 1997; Hawkins et al. 1998).

In conclusion, the Ulysses spacecraft detected the high latitude particle layer during both the inbound and outbound trajectory confirming previous reports about a large scale region of energetic particles and that Ulysses visited three distinct particle regimes during its flyby of the planet: the radiation belts in the inner magnetsophere, the magnetodisc, and the high latitude energetic particle layer in the outer magnetosphere. In this paper we provided several tests for supporting the hypothesis that certain Ulysses/HI-SCALE cross-B field anisotropies observed between the magnetodisc-related ion events in the middle magnetosphere and others within the outer magnetosphere outbound were mostly due to intensity gradients allowing the appearance of a third major energetic ion region in the Jovian magnetosphere.

A reexamination of the previous literature concerning the energetic particles populating the Jovian magnetosphere suggests that the discovery of the outer magnetosphere particle layer explains previous comments on some "strange" observations, which were reported as not consistent with the magnetodisc model. This problem was posed from the authors who noticed that Pioneer particle intensity levels were in the high latitude outer Jovian magnetosphere above those observed in the magnetosdisc (Kivelson, 1976; Kennel and Coroniti, 1979).

Furthermore, in this study we found strong evidence for the existence of a thin boundary layer of energetic particles close to the duskside south magnetopause.

The discovery of the energetic particle layer in the outer magnetosphere is very important, since it changes our phenomenological picture of the Jovian magnetosphere, with the revelation of a third major particle region of the Jovian magnetosphere, beside the radiation belt and the magnetodisc. In addition, this layer in the dusk south magnetosphere coincides with the region, where the ~15-20 min and ~40min quasi-periodic ion emissions coming from the poles were observed (Anagnostopoulos et al., 2001; Karanikola et al., 2004) and it appears as an important factor of the Jovian dynamics; this region constitutes a second reservoir of energetic particles, beside the magnetosdisk, for the particle leaking in the magnetosheath and the region upstream from the bow shock. The role of the high latitude outer magnetosphere ion region also reveals its significance from the fact that ~40min quasi-periodicities were observed up to at least ~1 AU from the planet in both the south and the north heliosphere, during the first (year 1992) and the second (years 2003-2004) trip of Ulysses in the environment of the Jovian magnetosphere (Marhavilas et al., 2001; Anagnostopoulos et al., 2009).

There is still a question whether the outer magnetosphere particle layer that discovered by Ulysses is a temporal particle concentration observed during the Ulysses and Pioneer 11 missions (these spacecraft visited the high latitude magnetosphere) or it is a steady third major particle region of the Jovian magnetosphere [Cowley et al., 1996]. In a future paper we will present comparative observations from Ulysses, Voyager 1 and 2, and Pioneer 11 and 12, which reveal that a similar structure was detected within the Jovian magnetosphere for all these five missions. Furthermore, in another paper ready to be submitted for publication we provide strong evidence that a similar high latitude region is also evident in relativistic electron data.

**Acknowledgments**

The authors appreciate the access to magnetic field data (courtesy of Ulysses VHM/FGM magnetometer team) and to relativistic electron measurements (courtesy of Ulysses COSPIN experiment team).

# Tables

**Table 1**

The table presents the evaluated gyroradius and anisotropy (according to Eq. 1) at $\sim 90^0$, for the mean energy of the proton channels W1 and W2, and the high energy heavy ion channel Z3 (assumed Z=5 for the gyroradius calculation), in each of the six measurements displayed in Fig. 8. It is evident that the anisotropy magnitude A is ordering according to the gyroradius value $R_g$, with increasing A magnitude as $R_g$ increases. This ordering is consistent, according to the previous analysis (Eq. 4) with cross-field anisotropy caused by ion intensity gradient, and not by the Compton-Getting effect.

|  | **W1** | **W2** | **Z3** |
|---|---|---|---|
| Energy range (MeV) | 0.5-0.95 | 0.95-1.6 | E>2.5 Z>5 |
| Gyroradius ($R_J$) | 0.17 | 0.22 | 1.60 (Z=5) or 5.11 ($O^+$, Z=16) |
| Anisotropy (d42 1608-1609 UT) | 0.33 | 0.54 | 0.57 |
| Anisotropy (d43 0100-0110 UT) | 0.20 | 0.29 | 0.34 |

# Figures

**Fig. 1.** The drawing shows the sector arrangement of the CA60 telescope of ULYSSES/HI-SCALE instrument in respect to the ecliptic and the Ulysses' trajectory during its inbound trajectory to Jupiter .

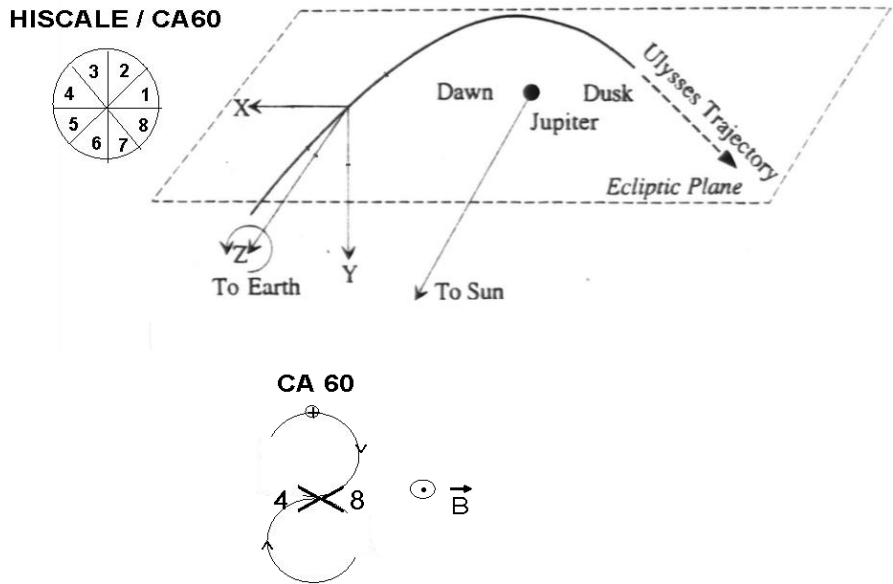

**Fig. 2.** A schematic representation of the magnetodisc energetic particle layer (plasma sheet) and the high latitude particle layer, in the XY plane during the Ulysses inbound trajectory days 33-39, 1992. The intensity anisotropy observed in oppositely looking sectors, that count particles perpendicular to the magnetic field, show the gradient direction and reveals the approach of the spacecraft to one of the two layers

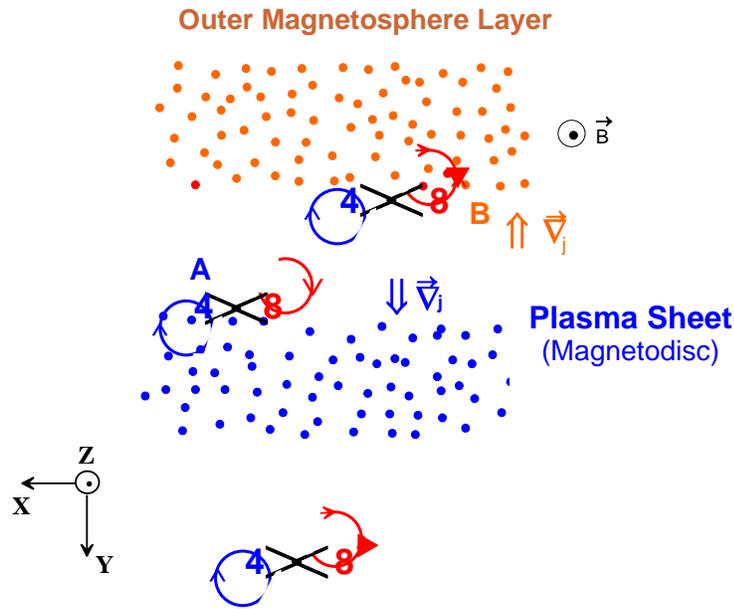

**Fig. 3.** Intensity of heavy ions (Z3) obtained by the CA60 telescope of the HI-SCALE instrument (panel a), intensities of heavy ions (Z3) in oppositely looking sectors that count perpendicular to the magnetic field (panel b), the anisotropy ratio of the intensities shown in panel b (panel c), the relative intensity gradient (panel d) and the intensity and the $B_\varphi$ component of the magnetic field (panels e, f). The positive values A, Δ, H observed in the anisotropy ratio show the intensity gradient toward the plasma sheet, while the negative values B, Γ, E and Z show intensity gradient toward the north verifying the existence of the high latitude particle layer

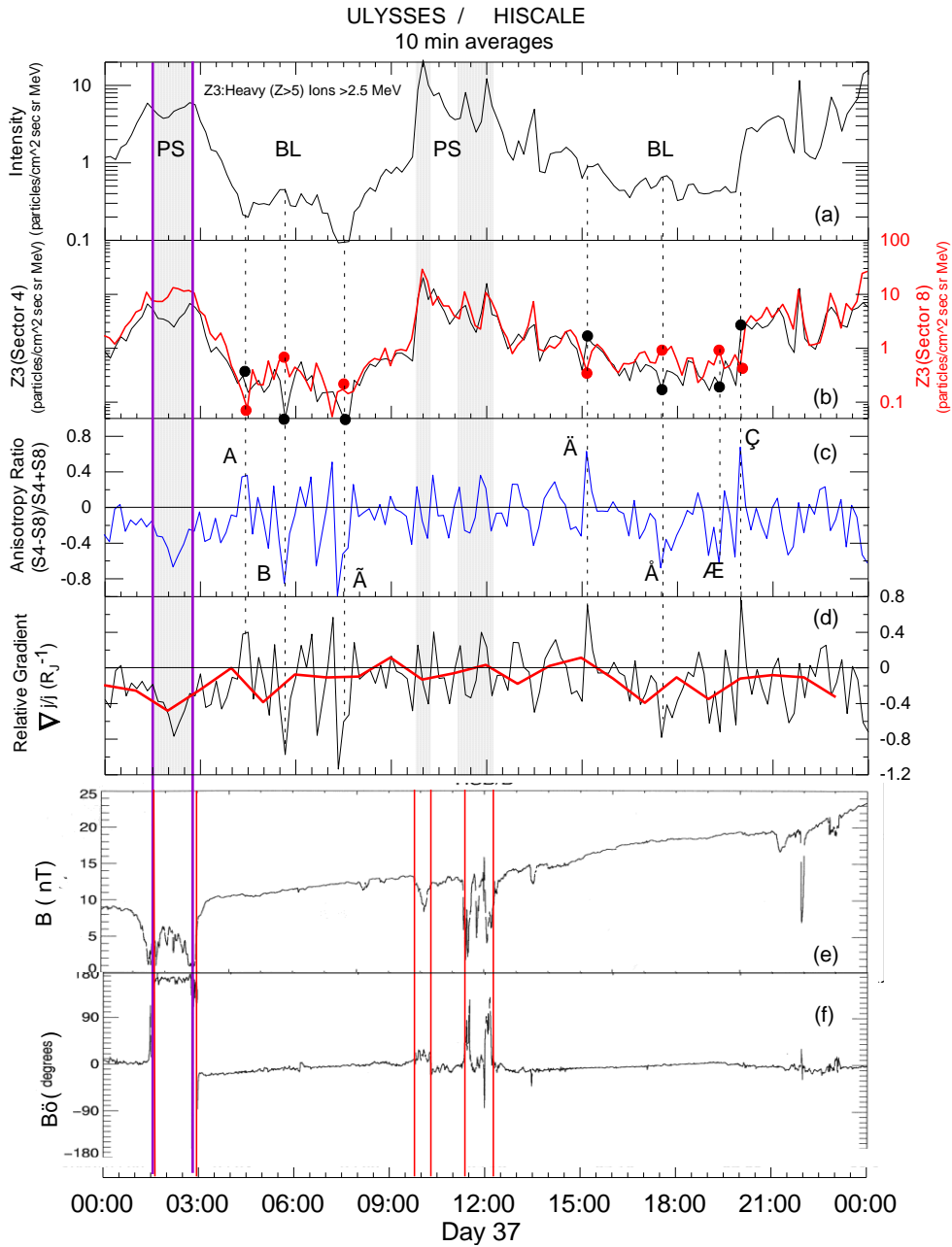

**Fig. 4.** As in the previous figure but for the time interval 1500-2400 UT, day 36, 1992. The positive peak A of the anisotropy index show intensity gradient toward a direction south of the spacecraft (plasma sheet), while the negative peaks B, Γ and Δ show intensity gradient toward the north (high latitude particle layer).

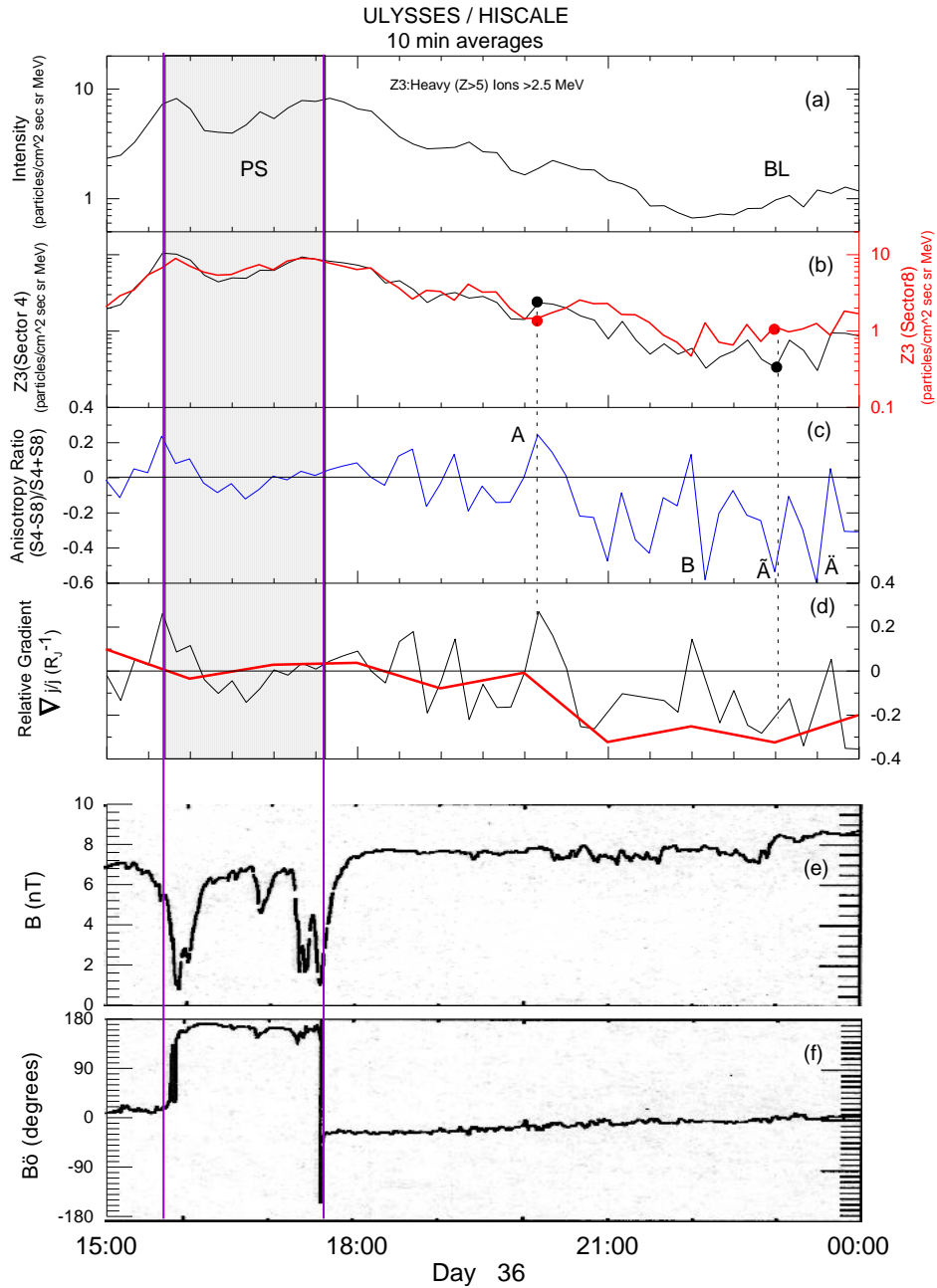

**Fig. 5.** As in Figures 3 and 4 but for measurements of protons 0.5-0.95 MeV (channel W1 of HI-SCALE/CA60 telescope), for the time interval 0000 UT, d 40 until 0600 UT, d 41, 1992, during Ulysses' removal from Jupiter, in the south hemisphere. The negative values of the anisotropy index which are observed simultaneously with the increase of the particle intensity (intervals: (i) ~0300-1000 and ~1300 – 2100 UT, d40, and (ii) ~0000-0600 UT, d41), show an intensity gradient in a direction toward the north (where is located the plasma sheet). Between these intervals, positive values of the anisotropy ratio are observed, showing an intensity gradient in a direction toward the south (where is located the high latitude particle layer)

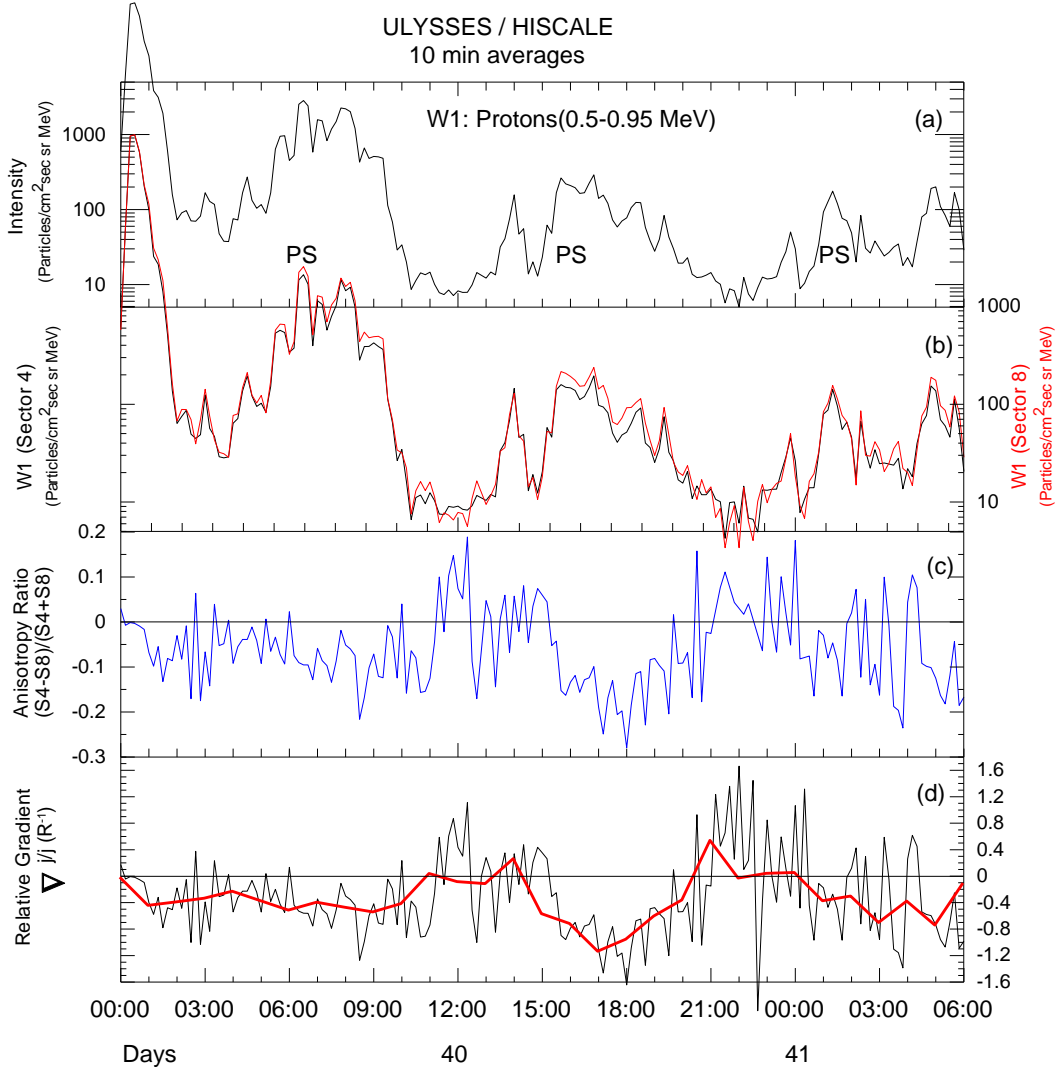

**Fig. 6.** Intensity profiles of (a) heavy ions obtained by the Z3 channel of the CA60/HI-SCALE detector, (b) sectors S4 and S8 (of the same channel) that count perpendicularly to the magnetic field, and (c) the profile of the anisotropy-index using the intensities of these sectors. An increase of the anisotropy is observed whenever Ulysses was approaching / traversing the magnetopause.

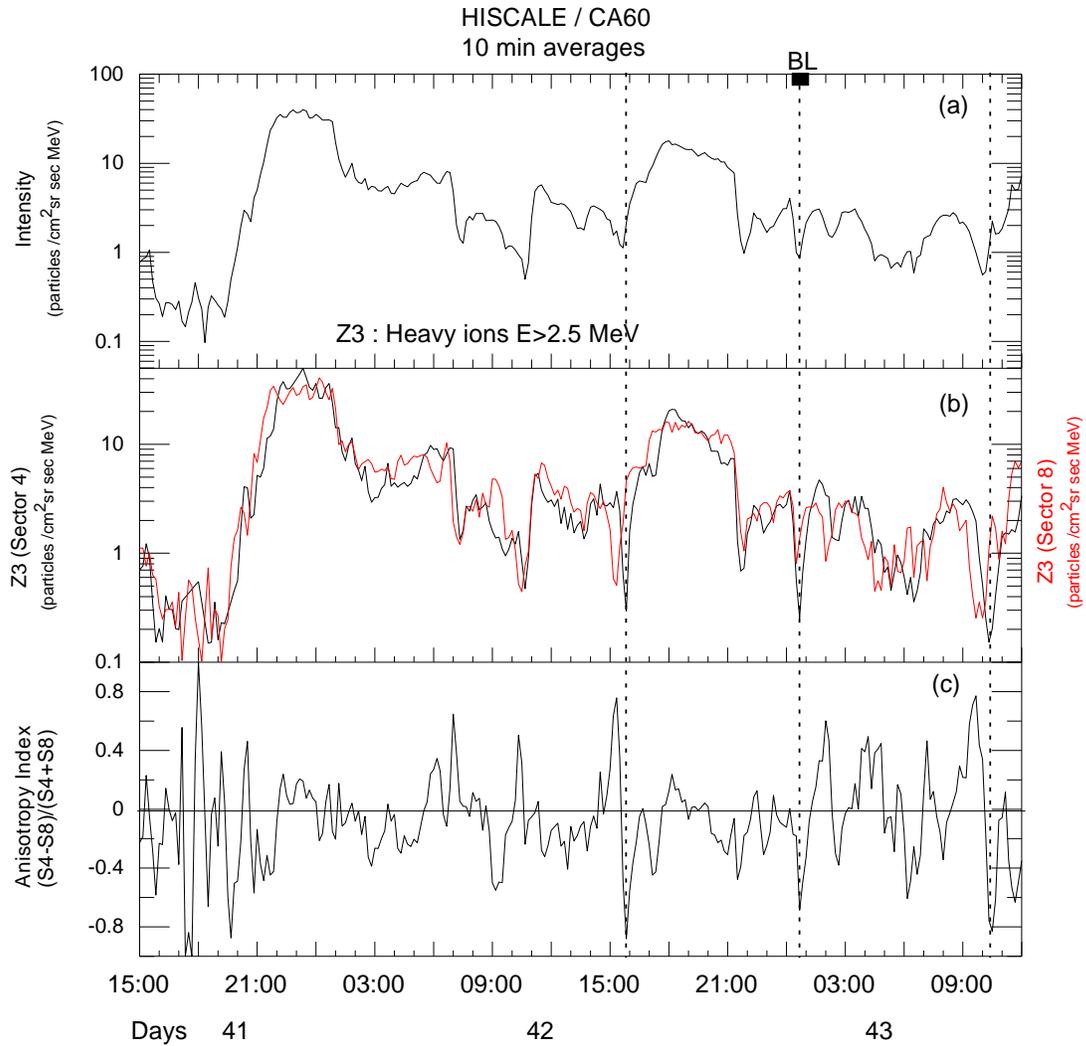

**Fig. 7.** The figure presents an anisotropy increase, perpendicularly to the magnetic field, in (i) Z3 heavy-ions' (E>2.5 MeV) profiles of CA60/HI-SCALE [drawing-set (a)], and (ii) H7S relativistic-electrons' (E>3MeV) profiles of HET/COSPIN [drawing-set (b)], whenever Ulysses was approaching/traversing the magnetopause. The direction of the electrons' anisotropy-vector is opposite with the heavy-ions' one (because the electrons gyrate oppositely to ions), indicating that the observed anisotropy is caused by the intensity gradient and not by the Compton-Getting effect, which generates intensity-increase in the same direction for both ions and electrons

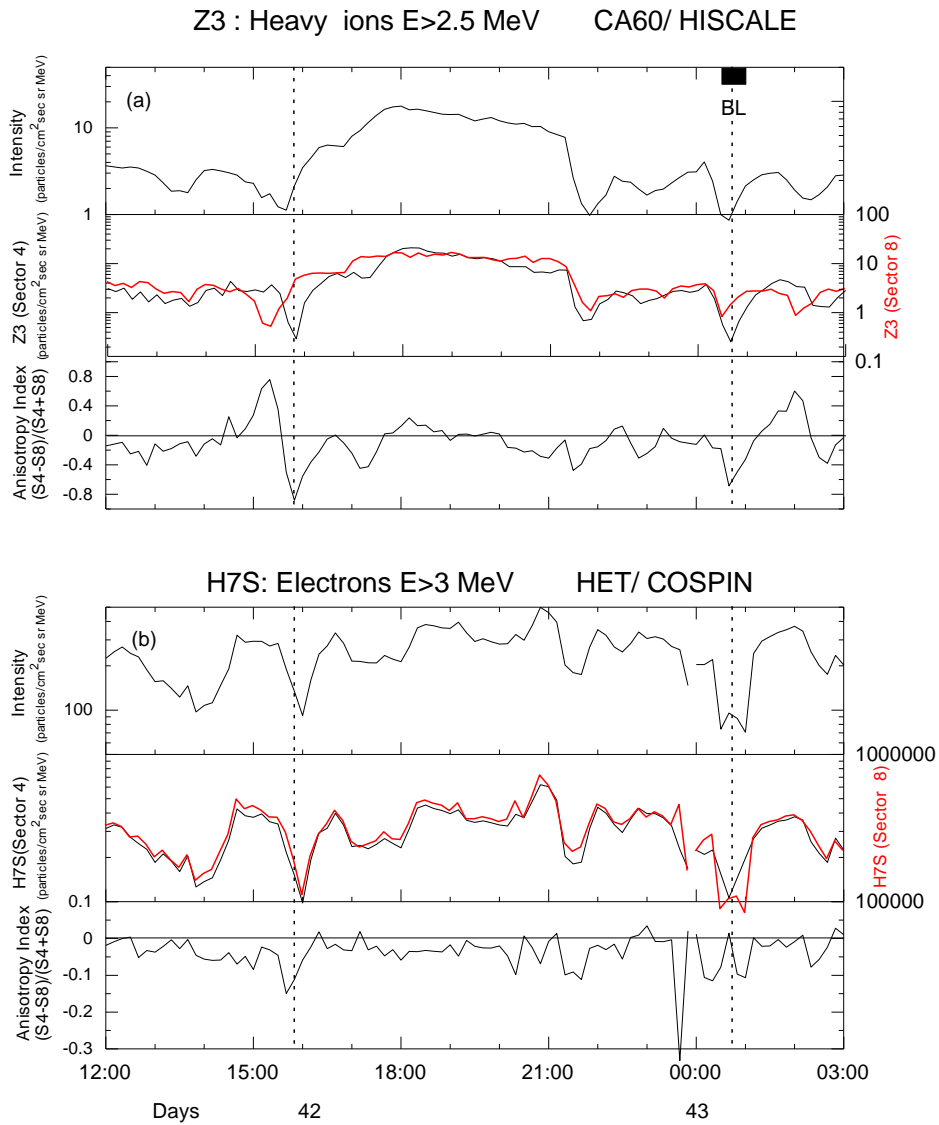

**Fig. 8.** Pitch-angle distributions of heavy ions Z3 (Z>5, E>2.5 MeV) and protons W1 (0.5-0.95 MeV) and W2 (0.95- 1.6 MeV) of the CA60 /HI-SCALE detector. Anisotropy at $90^0$ pitch-angle is observed that is more intense for the heavy ions and the higher energy protons, supporting the opinion that this is caused due to intensity gradient that depends on the particle's species and energy.

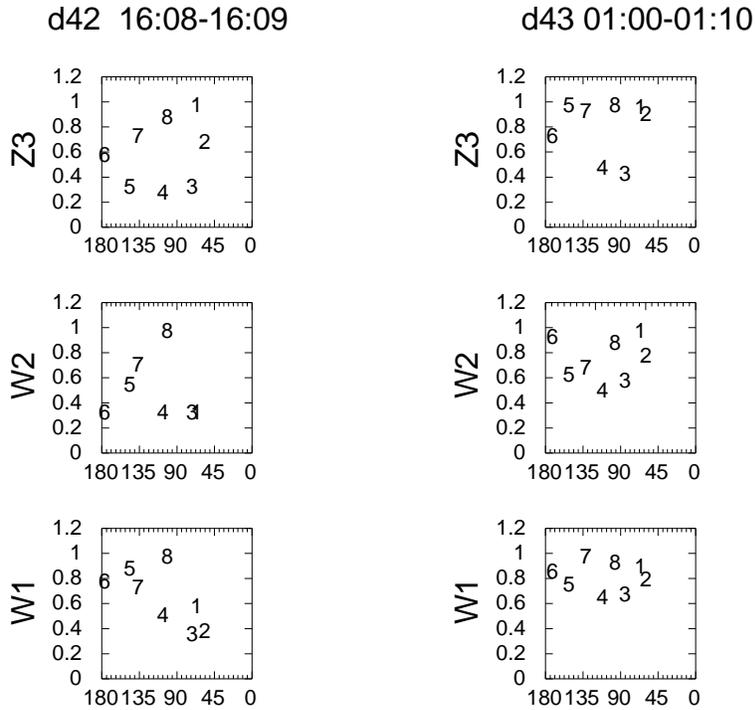

**Fig. 9.** As in Figure 5 but for the time interval 0900-2400 UT, d42, 1992. The change of the intensity gradient direction during the time intervals 1515-1645 UT (peaks A, B and Γ of the anisotropy index) and 2115-2200 UT (peaks Δ and E), when Ulysses was approaching the magnetopause, constitute a strong indication of the existence of a small layer of energetic particles close to the magnetopause

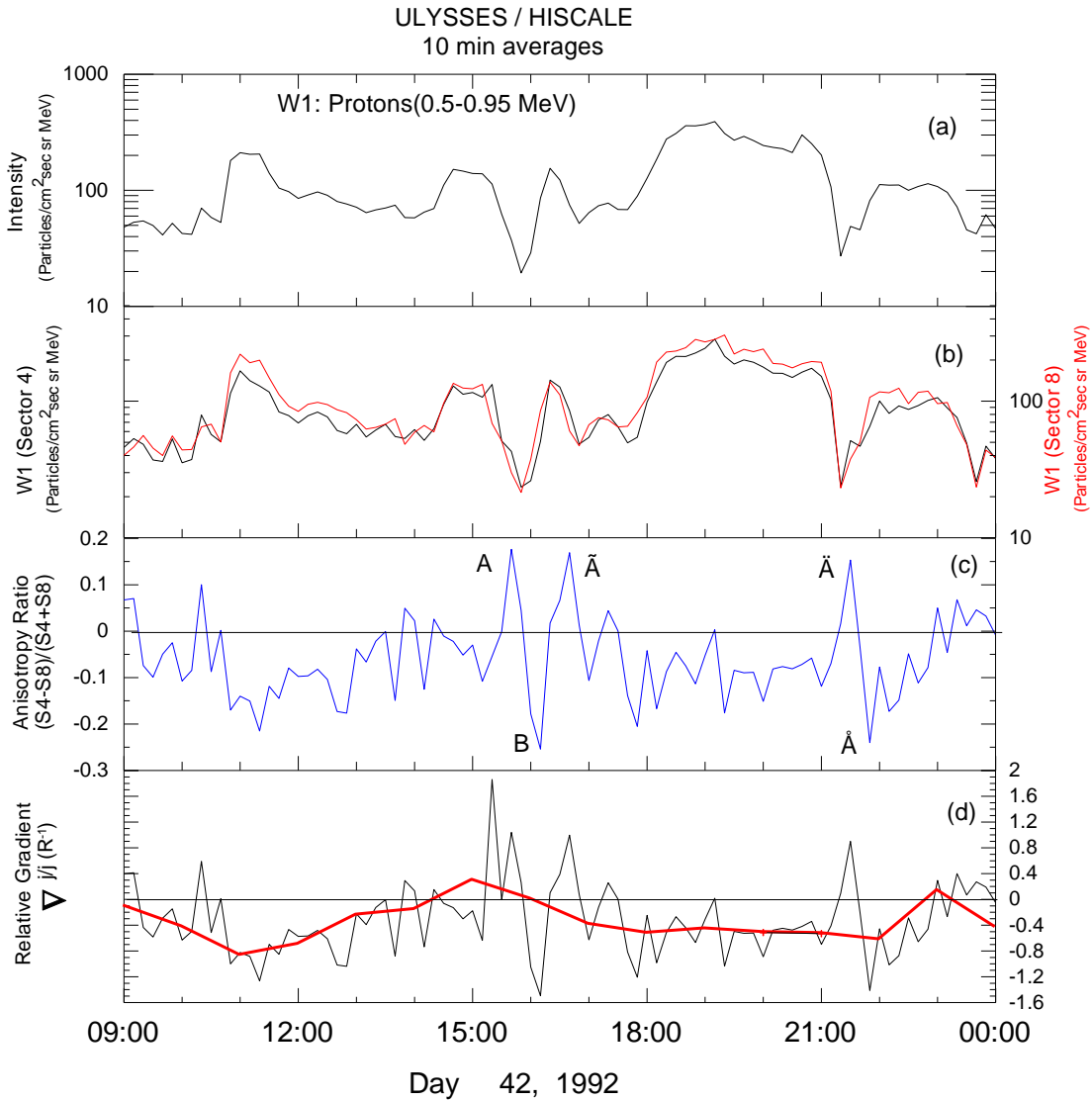

**Fig. 10.** A schematic representation of the W1 (0.5-0.95) MeV proton distribution, which is based on the Ulysses' observations, during its outbound trajectory of Jupiter, when the s/c moved at high latitudes in the duskside Jovian magnetosphere

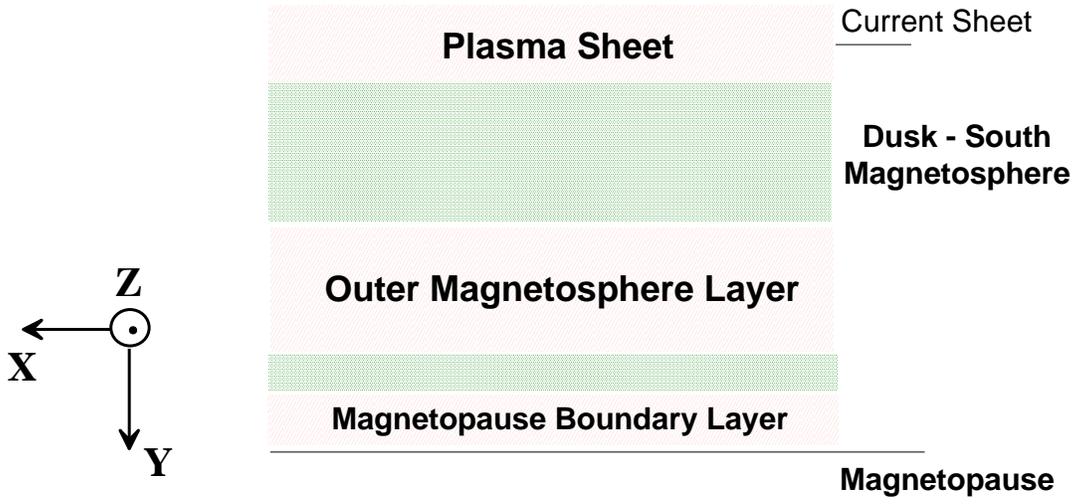

**Fig. 11.** A model of the Jovian magnetic field (Connerney et al., 1981) and the Ulysses' trajectory in magnetic coordinates. The red arrows indicate the time intervals during which an intensity gradient toward the high latitudes was observed, that verifies the existence of the recently discovered layer of energetic particles

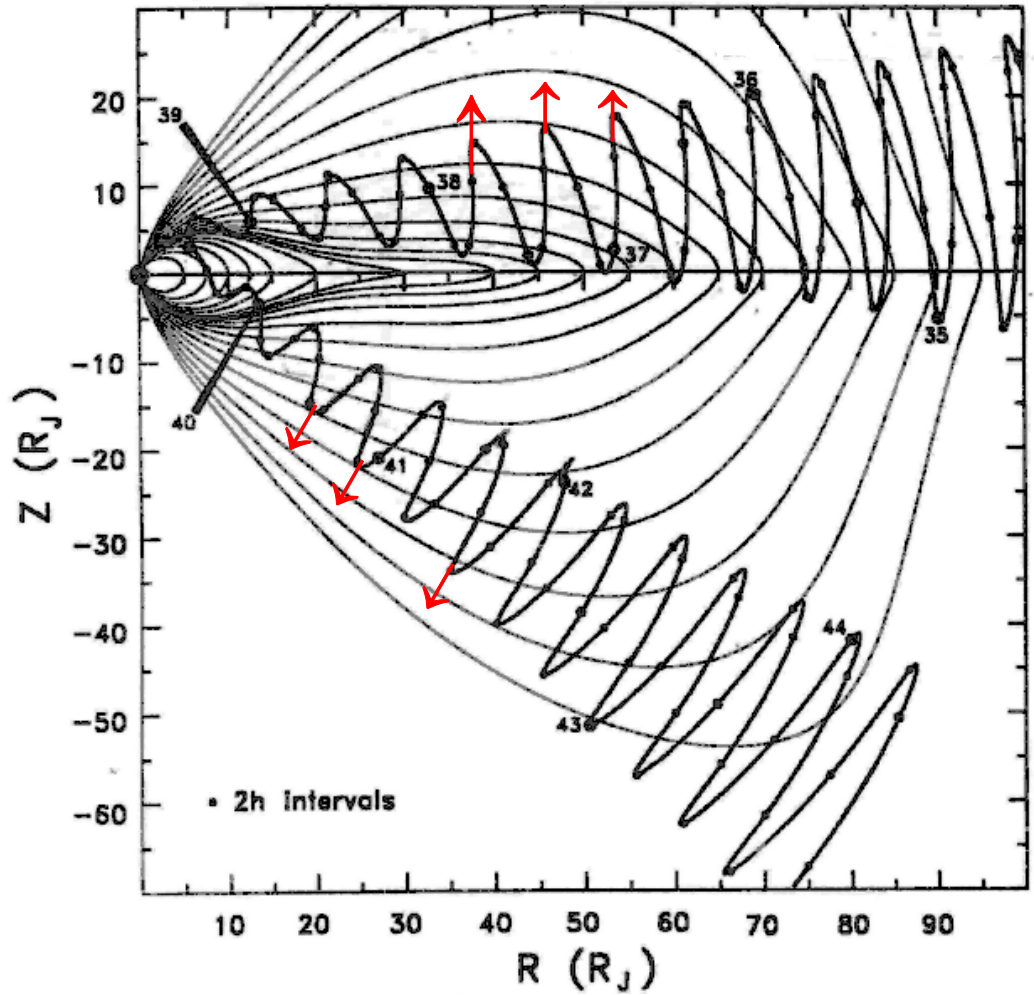